\documentclass[useAMS,usenatbib,usegraphicx]{mn2e}

\usepackage{amsmath}
\usepackage{url}

\def\fermi{{\it Fermi\/}}

\def\eg{{\it e.g.,}}
\def\ie{{\it i.e.,}}
\newcommand\lsim{\mathrel{\rlap{\lower4pt\hbox{\hskip1pt$\sim$}}
        \raise1pt\hbox{$<$}}}
\newcommand\gsim{\mathrel{\rlap{\lower4pt\hbox{\hskip1pt$\sim$}}
        \raise1pt\hbox{$>$}}}

\title[Annihilating dark matter or noise?]{Annihilating dark matter or noise?:
A statistical examination of the \fermi~GeV excess around the Galactic Centre}
\author[N. Mirabal]{N. Mirabal$^{1}$\thanks{E-mail:
nestor.r.mirabalbarrios@nasa.gov}\\
$^{1}$Astrophysics Science Division, NASA Goddard Space Flight Center, 
Greenbelt, MD 20771, USA
}

\begin{document}

\date{}

\pagerange{\pageref{firstpage}--\pageref{lastpage}} \pubyear{2014}

\maketitle

\label{firstpage}

\begin{abstract}
Excess \fermi~GeV emission around the 
Galactic Centre  has been  interpreted as a possible signature of
annihilating dark matter. Here we analyse three aspects of this
claim: its spectral cutoff, the correlation between the spectral shape
of the purported signal
 and known noise components, and its brightness profile.  
Experimentally, the correlations that exist 
between the GeV excess and known sources of
noise make it difficult to conclude that a 
dark matter signal claim can be made 
without independent direct detection confirmation. 
As a possible way forward, we introduce three criteria that could potentially 
help to validate a dark matter annihilation signal in gamma rays. 
\end{abstract}


\section{Introduction}
Distinguishing signal from noise is an everyday problem for 
experimental scientists. Examples of noise mimicking a potential signal 
will be familiar to everyone working in astrophysics. 
The {\sc Bicep2} finding of a $B$-mode polarization pattern is just 
the latest public example where residual
foregrounds
confused the sought signal 
\citep{bicep2}.
In retrospect, the early {\sc Bicep2} claim resembled a 
dust polarization signature but it was difficult to ascertain  
at the time of the original claim
\citep{flauger}. 

Noise subtraction can be especially daunting in situations when the 
signal has to be extracted simultaneously 
from the same region of the
sky and without a well-characterised noise model from 
neighbouring frequencies. 
Such is the case when attempting a detection of annihilating 
dark matter  at the Galactic 
Centre. Over the past few years, a number of groups
have found possible excess \fermi~GeV emission around 
the Galactic Centre \citep{gh,hl,boyarsky,
abazajian,daylan}. This result has opened the door to an
avalanche of publications about dark matter models and 
interpretations. While it is wonderful to see practical applications of 
theoretical dark matter models and a possible solution to one of the
hardest problems in science today, we might be over-interpreting the
evidence.

Rather than trying to delve into another discussion about 
the origin of the GeV excess, we will try to reduce our work to a more
basic proposition: does the GeV excess 
constitute sufficient evidence of a dark matter signal or reflects the
limitations of the current noise model?. Or more specifically, 
are the GeV excess and known sources
of noise at the Galactic Centre uncorrelated?. We assume the perspective 
of an experimentalist
perched on a laboratory stool pondering
recent findings. 

For practical purposes, we consider a situation in which 
dark matter annihilation represents the sought signal. Every other 
non-related gamma-ray 
contribution in the region of interest represents noise. We thus interpret 
noise in its broadest sense with four principal 
known noise components: $\pi^{0}$ decay 
\citep{ackermann2012}, inverse Compton \citep{ackermann2012}, bremsstrahlung 
\citep{ackermann2012}, and   
unresolved point sources \eg~ millisecond pulsars (MSPs) \citep{abazajian,
mirabal2, yuan,calore}. For a more complete picture, one 
should also account for all the  
uncertainties/unknowns that can potentially contribute to the noise model. 
Figure \ref{figure1} shows a schematic view  of the noise components 
considered here. 

In this brief note, the goal is to conduct 
a statistical assessment of the GeV excess. For this purpose we consider 
three specific aspects of the GeV excess:
its spectral cutoff (Section \ref{section2}), 
the correlation between the spectral shape of the purported signal
 and the noise components (Section \ref{section3}), and its 
brightness profile (Section \ref{section4}). Finally, some conclusions
and selection criteria are presented in Section \ref{section5}.

\section{Spectral cutoff}
\label{section2}
One of the most surprising facts about the 
GeV excess is that it displays
a spectral cutoff that is aligned with the cutoff 
observed in the spectrum of \fermi~MSPs \citep{hl,abazajian}.
A few years ago \citet{baltz} anticipated 
that the spectral similarities between 
annihilating weakly interacting massive particles (WIMPs) and  
gamma-ray pulsars would be the most problematic 
obstacle to proving a dark mater astrophysical 
signal. Their reasoning was rather straightforward: 
pulsar emission and dark matter annihilation are predicted to share
similar spectral signatures with sharp cutoffs.
However, the idea was introduced in the context of dark matter subhalos and
preceded all the extraordinary 
{\it Fermi} discoveries \citep{2p}. 
The intricacies of the Galactic Centre did not even enter 
into the discussion. A placement of the GeV excess 
in the Galactic Centre adds an extra layer of complexity.

\begin{figure}
\hfil
\includegraphics[width=3.5in]{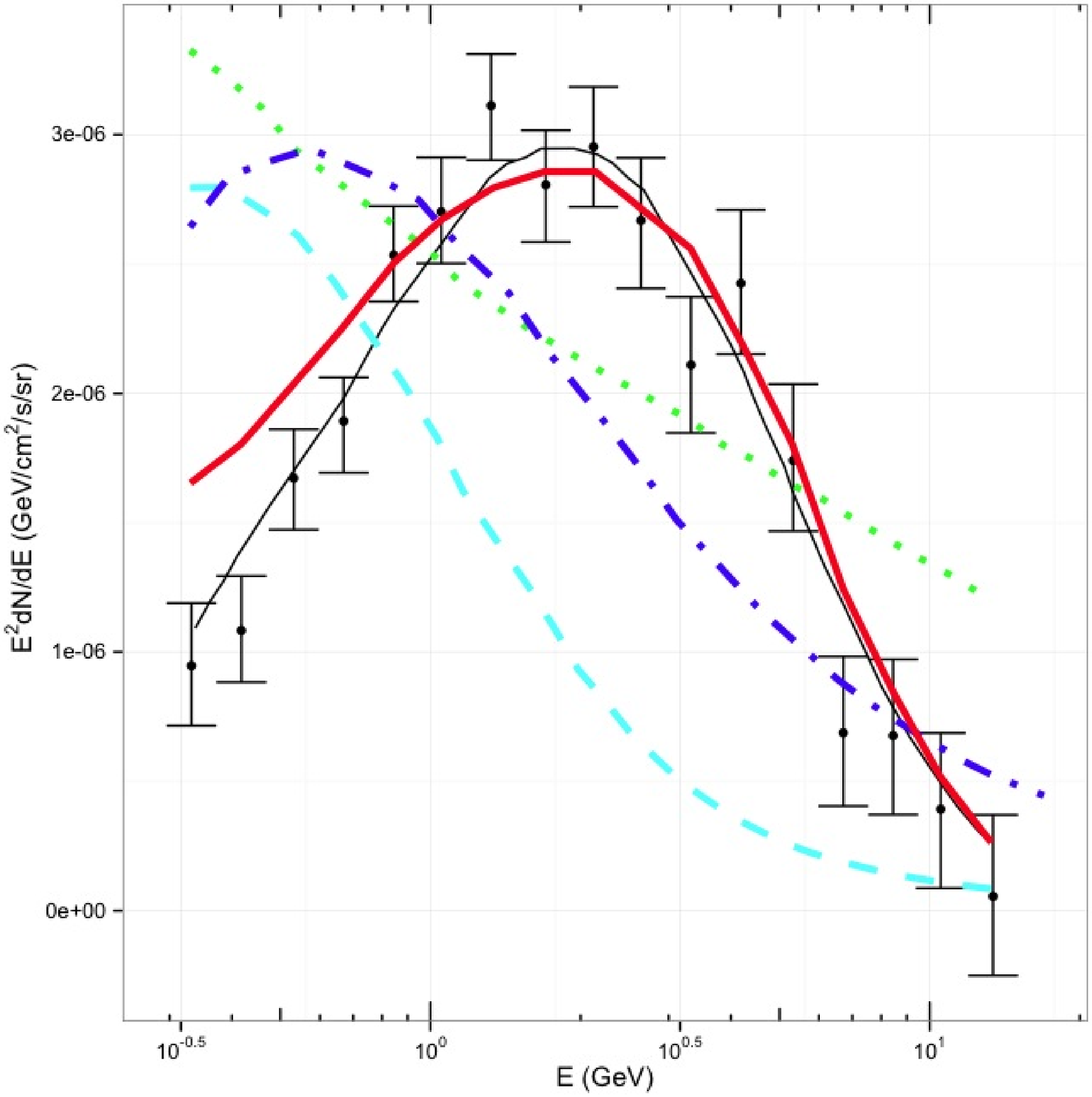}
\hfil
\caption{
The spectrum of the GeV excess for an NFW profile with  $\gamma=1.26$ 
from the analysis by 
\citet{daylan}. Shown in the black solid
line is the spectrum for a 35 GeV dark matter particle 
annihilating to $b\bar{b}$ with a cross section 
$\sigma v = 1.7\times 10^{-26}$ cm$^3$ s$^{-1}$  \citep{daylan}. 
The noise is split into four components: 
$\pi^{0}$ decay (blue, dash-dotted), 
inverse Compton (green, dotted), bremsstrahlung 
(cyan,dashed) all taken from \citet{ackermann2012}, 
and the average unresolved MSP spectrum (thick solid red)
from \citet{calore}. All components have been
scaled in flux for clarity.
}
\label{figure1}
\end{figure}

Figure \ref{figure1} clearly illustrates the remarkable 
match between the GeV excess and the unresolved MSP noise component.   
Shown are the GeV excess data set and a dark matter
template from the analysis performed by \citet{daylan}. 
For comparison, we also plot the average contribution 
from unresolved MSPs
in the Galactic Centre derived by \citet{calore}, scaled in flux. 
It is easy to see the conspicuous alignment of the cutoff at 
energies greater than 2 GeV. 

The 100 MeV-1 TeV energy range covered by {\it Fermi}-LAT
is the preferred scale for 
a myriad of viable dark matter particle models in
the literature.  Sticking 
with some of the most popular, models stretch from the WIMPless in the
MeV range \citep{feng,albert} to WIMPs in the GeV to TeV range 
\citep{jungman}.
Our first question is obvious: what is the probability that out of 
all well-motivated dark matter candidates the one found 
shares the same spectral cutoff with a known noise component, 
namely unresolved MSPs?.

The problem would be very difficult to solve if MSP spectral cutoffs
were distributed over a very wide energy range. However, the observed 
MSP spectral cutoffs actually cluster in a fairly narrow energy band  
0.84 GeV $\lsim E_{\rm cut} \lsim$ 5.4 GeV \citep{2p}. 
To  estimate the probability of a chance alignment between dark matter 
annihilation and 
a  noise component, one can solve  a  version of 
the classical birthday problem \citep{vonmises}.
The dark matter candidate must match the MSP cutoff 
within a very narrow energy bin width $\Delta E \sim 5$ GeV. Accordingly, 
there are 200 possible energy bins 
between 100 MeV and 1 TeV where this might have happened.  
Then the probability of having the same energy cutoff $P(same)$
by chance is rather small 

\begin{equation}
P(same) = 1 - P(different) =  1 - \frac{199}{200} = 5 \times 10^{-3}.
\end{equation} 

A chance alignment is even less likely if one chooses to include 
sterile neutrinos and axions as possible dark matter candidates over
a much wider energy range. 

\section{Correlation between noise and signal}
\label{section3}
Next, we extend our analysis by 
considering the GeV excess over the entire {\it Fermi} range. 
\citet{abazajian} noticed that the MSP noise component tends to be 
softer than the GeV excess below 0.8 GeV (see Figure \ref{figure1}). 
One could suppose that over-subtraction at energies below 1 GeV
of any of the noise components 
rendered in Figure \ref{figure1} is the cause
for the slight mismatch. But let us rather pose the following question: 
is there a correlation between the spectral shape of
the GeV excess and the spectral shape of individual 
noise components?. 

For this exercise, we take the 17 spectral points from 
\citet{daylan} and find their corresponding E$^{2}$ dN/dE intercepts in the
\citet{ackermann2012} and \citet{calore} curves. 
We then perform a Spearman correlation analysis and find the highest
Spearman's rank correlation coefficient
$\rho =0.973$ with a  $p$-value of $p_{\rm s}$ = $5.8 \times 10^{-11}$ for
unresolved MSPs,
which suggests that a real correlation exists between the purported signal
and a known source of noise. 

Despite a near perfect spectral match, 
it has been argued that the predicted flux level of the MSP noise component 
is considerable lower than the observed GeV excess \citep{cholis}. 
In order to study the influence of the flux normalization, we run 
a Kolmogorov-Smirnov (KS) test between
the \citet{daylan} points
and the \citet{calore} values scaled in flux  ($15\times$), which  
finds consistency with 
the samples having been drawn from the same parent distribution ($P=0.98$).

Taken together, these significant correlations 
between the spectral shape of the GeV excess and the
spectral shape of a known background could instead be a
symptom of problems with the normalization term   
in a very complex noise subtraction exercise \citep{strong,porter,casandjian,ackermann2012}.

\section{Brightness Profile}
\label{section4} 
\begin{figure}
\hfil
\includegraphics[width=3.5in]{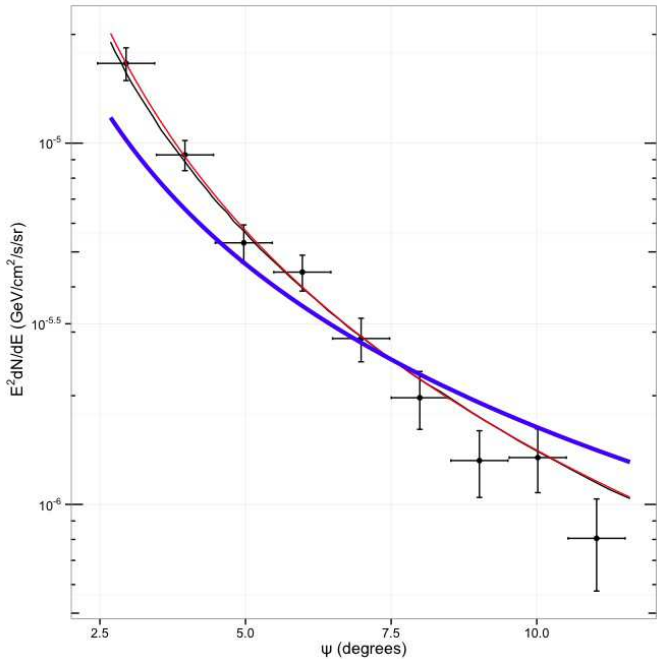}
\hfil
\caption{Flux points from the concentric ring analysis carried out by 
\citet{daylan}. The dark line shows the prediction for a generalized
NFW profile with $\gamma = 1.4$. In red (nearly indistinguishable) 
we show an $\psi^{-2}$ fit with no prior assumption about dark matter. 
The thick blue line shows the $\psi^{-1.5}$ radial distribution of LMXBs
in M31 \citep{abazajian}.}
\label{figure2}
\end{figure}

Finally, the third aspect of the GeV excess we would like
to discuss is its spatial extension. 
A  canonical Navarro-Frenk-White 
(NFW) density 
profile with $\gamma =1.4$ fits the brightness profile out to $\sim 12^{\circ}$
\citep{daylan}. 
Figure \ref{figure2} shows the brightness profile from \citet{daylan} and the
corresponding NFW profile. An NFW density profile is quite majestic 
\citep{nfw}, but
in reality an observer can only measure a projected brightness profile. For
Over small radial ranges, a generalized NFW profile 
asymptotes to a rather generic brightness fall that follows a single 
power-law  index. As an illustration, we show an 
inverse-square law in Figure \ref{figure2}
where one can see that the NFW fit from \citet{daylan} 
and the $\psi^{-2}$ power law with no prior assumption about dark matter
are nearly indistinguishable. 

Our third question takes form: 
are these radial trends unique 
to annihilating dark matter?.  
Given the complexities of our own Galactic Centre and the extent of the 
GeV excess,  
we look for guidance in our nearest neighbour, M31. 
Andromeda is the nearest spiral galaxy 
similar to our own Milky Way and offers a unique 
line of sight encompassing both its
bulge and galactic centre. Similar to the properties of the Milky Way, 
M31's stellar halo follows a power-law component
with index $-2.2 \pm 0.2$ \citep{gilbert}. 
Further, \citet{abazajian} first noted that the  
low-mass X-ray binaries (LMXBs) population in M31 follows a
power-law index of $-1.5 \pm 0.2$ (see Figure \ref{figure2}).
If LMXBs are the progenitors of MSPs \citep{grindlay,kulkarni}, 
it is reasonable to expect that the unresolved
MSP contribution will follow a similar spatial distribution.  

Even if one dismisses known point sources as
the perpetrators, M31 shows unresolved diffuse X-ray emission  
that extends to $8^{\prime}$ from its galactic centre  
\citep{li,bogdan}, which at the distance of
our own Galactic Centre would fill its surrounding 
$\sim 13^{\circ}$. This extension approximately matches 
the observed width of the GeV excess.
There are related hints of unresolved diffuse X-ray emission in
our own Galactic Centre that might indicate a population
of uncharted high-energy sources or serious shortcomings in
our understanding of the interstellar gas \citep{yusef}.

Although not causally connected, 
these are spatial noise templates that could account for both
the extent and profile of the GeV excess without direct dark matter 
involvement.

\section{Conclusion and Future Work}
\label{section5}
We have considered three aspects of the GeV excess at the Centre of our 
Galaxy. As an experimentalist, the correlation that exists between 
individual noise components and the purported GeV excess  
cannot convincingly establish that it is 
an actual signal under the broad noise definition introduced here. 
Therefore, we conclude that the discovery criteria
required to make a dark matter claim 
has not yet been met. 

Using anomalies in the Galactic Centre to help pinpoint similar signatures 
in other sections of the Galaxy including dwarf galaxies and dark matter
subhalos appears to be a sensible approach, but it might not be
the most efficient route to take in every instance. 
A large variety of alternative searches that are conducted 
routinely by the {\it Fermi} collaboration 
and other groups can be used in lieu of the Galactic
Centre approach \citep{lines,dwarfs}. 

Regardless of the final verdict on the GeV excess, 
it is clear that a more comprehensive discussion about what constitutes 
evidence for an 
annihilating dark matter signal in our Galaxy is needed, as other 
claims might appear again in the future.  
For an opening proposal, we suggest imposing the following three criteria
before making any claim:
\begin{enumerate}
\item There should not be a near perfect energy alignment between the 
 dark matter spectral break and the cutoffs of any well-known noise components.
\item There should not be a correlation between 
the spectral shape of the dark matter signal and the spectral shape of 
known sources of noise.
\item There should be a spatially extended brightness profile. If
the generic power-law index corresponds to $m = 2.0 \pm 0.2$,
then we can borrow the traditional criterion
for discovery in particle physics
by requiring that an annihilating
dark matter brightness profile be at least $5\sigma$ away
and falls off as $\psi^{-m}$, with
\ie~$m \lsim 1$ or $m \gsim 3$ for a distinct signature.
\end{enumerate}

In our humble opinion, any dark matter signal worth pursuing in earnest 
should meet at least two out of the three 
criteria. Otherwise, direct detection is required to confirm any 
claim. With such tentative results, it might be more practical 
to devote theoretical efforts to other unsolved problems in astrophysics 
until 
a signal from  direct detection experiments is found and then and only then 
return to the {\it Fermi} archives for final astrophysical 
confirmation. 

Meanwhile, 
the appearance of tensions with other observations should 
be used discard claims that do not meet the 2/3 standard. In the 
case of the GeV excess, we are starting to reach the limits needed to 
evaluate possible tensions with 
dwarf galaxies observations \citep{geringer,anderson}. 

As it stands now, 
one might invoke a vast cosmic conspiracy at work in
the GeV excess (it is plausible)
that makes dark matter annihilation nearly indistinguishable from known 
sources of noise. Or maybe one should
pursue a possibly more banal explanation. To be fair, it is important to 
once again remark
that the proposed criteria 
discussed here do not rule out   
a dark matter origin for the GeV excess 
(only direct detection
limits can do that). In closing, we leave it to the adopters of these 
criteria
(if any) to reach their own conclusions and expand the list.

\section*{Acknowledgements}
I acknowledge support from the Spanish taxpayers 
through a Ram\'on y Cajal fellowship during the early stages of this work.
I thank Jos\'e Luis Contreras for helpful comments. 
I would like to acknowledge all the cooks at 
the ramen shop inside the Kanayama Station 
in Nagoya, where the bulk of the calculations were completed
during the Fifth International \fermi~ Symposium.   
This research was supported by a senior 
appointment to the NASA Postdoctoral Program
at the Goddard Space Flight Center, administered by Oak 
Ridge Associated Universities through a contract with NASA.

\label{lastpage}
\end{document}